\documentclass[acmsmall]{acmart}
\graphicspath{ {./fig/} }
\usepackage{pdfpages}

\AtBeginDocument{%
  \providecommand\BibTeX{{%
    \normalfont B\kern-0.5em{\scshape i\kern-0.25em b}\kern-0.8em\TeX}}}

\setcopyright{acmcopyright}
\copyrightyear{2018}
\acmYear{2018}
\acmDOI{10.1145/1122445.1122456}

\acmConference[Preprint]{cscw}{CSCW}{2021}
\acmBooktitle{Woodstock '18: ACM Symposium on Neural Gaze Detection,
  June 03--05, 2018, Woodstock, NY}
\acmPrice{15.00}
\acmISBN{978-1-4503-XXXX-X/18/06}

\begin{document}

\title{Promoting Self-Efficacy Through an Effective Human-Powered Nonvisual Smartphone Task Assistant}


\author{Andr\'e Rodrigues}
\email{afrodrigues@fc.ul.pt}
\affiliation{%
  \institution{LASIGE, Faculdade de Ci\^{e}ncias, Universidade de Lisboa, Portugal}
}

\author{Andr\'e Santos}
\affiliation{%
  \institution{LASIGE, Faculdade de Ci\^{e}ncias, Universidade de Lisboa}
  \country{Portugal}}
\email{arbsantos@fc.ul.pt}

\author{Kyle Montague}
\affiliation{%
  \institution{NorSC, Northumbria University}
  \city{Newcastle upon Tyne, Tyne and Wear}
  \country{United Kingdom}
}
 \email{kyle.montague@northumbria.ac.uk}

\author{Tiago Guerreiro}
\affiliation{%
  \institution{LASIGE, Faculdade de Ci\^{e}ncias, Universidade de Lisboa}
  \country{Portugal}}
\email{tjvg@di.fc.ul.pt}

\renewcommand{\shortauthors}{A. Rodrigues, A. Santos, K. Montague, T. Guerreiro}

\begin{abstract}
  Accessibility assessments typically focus on determining a binary measurement of task performance success/failure;  and often neglect to acknowledge the nuances of those interactions. Although a large population of blind people find smartphone interactions possible, many experiences take a significant toll and can have a lasting negative impact on the individual and their willingness to step out of technological comfort zones. There is a need to assist and support individuals with the adoption and learning process of new tasks to mitigate these negative experiences. We contribute with a human-powered nonvisual task assistant for smartphones to provide pervasive assistance. We argue, in addition to success, one must carefully consider promoting and evaluating factors such as self-efficacy and the belief in one's own abilities to control and learn to use technology. In this paper, we show effective assistant positively affects self-efficacy when performing new tasks with smartphones, affects perceptions of accessibility and enables systemic task-based learning. 
\end{abstract}



\maketitle

\section{Introduction}
Mastering an interactive technology, as with any other skill, requires training. When users struggle with a new device or interface, the most important aspect to adoption and learning is not how fast they accomplish a specific task, but rather how confident they are to experiment, and continue experimenting, until they master it. As described by \citet{Bandura1998Personal}, \textit{“Belief in one owns ability to produce the desired effects is the foundation of action, without it there is little incentive to act”}. 
 This relevance of self-efficacy has been acknowledged, in limited instances \cite{Yi2003Predicting}, for its important role in the adoption and use of technology.
Touch-enabled smartphones and their multi-faceted applications were an example of a disruptive new technology that imposed adoption barriers to users, particularly the non-tech savvy or those with disabilities. For instance, despite accessibility services, blind people still face challenges with these devices, that in extreme cases prevent adoption, or the use of new apps and features.
Nowadays, smartphones have built-in accessibility services (i.e. screen reader) that enable access to the devices for blind people. Although many of the functionalities can now be accessed, mastering the device requires significant effort and it is not without its challenges \cite{Damaceno2018Mobile, Grussenmeyer2017Accessible,Leporini2012Interacting,Rodrigues2015Getting}.  Adding to the lack of available support, prior research has also highlighted the issues users face when not confident in their own ability to control the device \cite{Pal2017Agency}. The need of an available support network has been identified previously for its potential to improve user self-organized learning \cite{Rodrigues2017In-context,Rodrigues2015Getting, Rodrigues2019Understanding}. 
In this paper, we explore if we can provide effective systemic support to blind users with their smartphone challenges, and if that impacts the level of confidence, they show in using the device and applications. Further, we explore if increased self-efficacy affects users’ overall perceptions of accessibility. We set out to answer the following research questions:
\begin{enumerate}
    \item Can a human-powered nonvisual task assistant enable effective smartphone assistance for blind users? (RQ1)
    \item Does effective smartphone assistance affect users’ perception of self-efficacy related with their smartphone competence? (RQ2)
\end{enumerate}
We designed and developed RISA, a human-powered nonvisual task assistant for smartphones. With RISA, we conducted a comparative user study where 16 blind participants performed a set of tasks, in mainstream apps, with and without assistance. The tasks were created based on the contributions of sighted participants with no domain knowledge (i.e. no accessibility awareness of how blind users interact with mobile devices). Our results show RISA significantly improved participants task success rate (70\% against 10\%). RISA enabled users to learn with a task-based approach, which was reported as ideal for novice users and for first app interactions. We found RISA had a positive effect on individual’s reported self-efficacy and perceived app accessibility, suggesting future work can leverage pervasive assistance to improve performance and internal (i.e. self-efficacy) and external (e.g. app accessibility) perception of users’. 
We contribute with 1) a human-powered nonvisual task assistant that relies on non-expert (i.e. people with no accessibility knowledge) authors; 2) a comparative assessment of the effectiveness of a human-powered nonvisual task assistance; and 3) an understanding of the impact of pervasive indirect assistance on perceived self-efficacy.

\section{Related Work}
Prior work in smartphone accessibility for blind people has targeted issues discreetly, often isolating the task from the context it is performed in, focusing on performance metrics, such as the works in text-entry\cite{Azenkot2012Input, Guerreiro2015TabLETS,Guerreiro2008From, Nicolau2014B}, and gesture input \cite{Buzzi2017Analyzing, Kane2011Usable, McGookin2008Investigating, Oh2013Follow}. 
Others have proposed solutions at a systemic level that adapt how content is presented \cite{Rodrigues2017Improving, Ross2017Epidemiology, Zhang2017Interaction}.  However, so far previous work on system-wide solutions has been devoid of any assessment of performance or impact on users’ perceptions of self-efficacy. 

\subsection{Smartphone Support}

In the two major smartphone operating systems (Android and iOS), screen readers are available by default, and the respective platforms provide a set of lessons (Android) or practice sandbox (iOS) to teach how to perform gestures, and manipulate a variety of interfaces (e.g. lists, edit boxes, virtual keyboard). Previous studies have argued that these offerings are ineffective to support the learning process and device adoption \cite{Rodrigues2015Getting}. Current \textit{getting started} approaches are focused on the structure of interfaces, and syntax of gestures, rather than guiding users through a set of tasks they want to learn how to perform (i.e. how to perform a call) \cite{Rodrigues2017In-context}. Users learn in an artificial context, a set of gestures, that they are expected to be able to transpose to real app interfaces.
Beyond accessibility, it is common for applications to provide on-boarding tutorials and other forms of in-app assistance. Users are guided through the main features of an app when they first interact with it. These onboarding tutorials are often designed to mimic the interface visually (app screenshots), to provide hints to app functionalities through occluding overlays with subtle markings, or visual representations of gestures to be mimicked by the user to proceed. This design aesthetic limits their usefulness to blind and visually impaired people, as the nonvisual connection is broken. Currently, the expectation is for developers to create accessible applications, and support for everyone. However, the reality is that apps are not accessible, and support mechanisms are non-existent, or do not match users' expectations \cite{Guerreiro2019Mobile, Ross2018Examining}. 
 \citet{Wang2014EverTutor:} have explored how to overcome the limitations of assistive content created by developers, and proposed a solution that enables the creation of system-wide tutorials based on user demonstrations. The work was focused on contextual step-by-step guidance with visual representations of the gestures to be performed at each step. \citet{Rodrigues2019Understanding} investigated how to leverage blind and sighted people's contributions, with limited accessibility knowledge, to create tutorials through demonstration. The results of the \textit{Wizard of Oz} study shown that there is a gap between what is provided by tutorial authors and what is required by the tutorial users, and further support was needed from co-located experts. 
Currently, when facing a challenge, instead of relying on co-located help, blind people can also rely on online forums dedicated to smartphone accessibility \cite{AppleVis}. However, to leverage these forums and groups, people have to be able to describe their issue, and to transpose whatever solution is provided to their specific context \cite{Bigham2017Effects}. To provide in-context assistance, \citet{Rodrigues2017In-context} have explored the acceptance of a human-powered question and answer service. Unfortunately, the system was only explored with a single contributor, an expert in accessibility, that was co-located and fully aware of the user context. 
Providing in-context assistance that adequately responds to the user’s needs is a demanding task. Current solutions are not accessible or rely on expert contributors to provide support. Prior work has highlighted the importance of effective support, but so far has been unable to provide it. Furthermore, past discussions focus on accessibility and on the binary measure of success/failure (i.e. can or cannot) without considering the nuances and relevance of its impact on self-efficacy and the users willingness to re-engage when faced with barriers and challenges. 

\subsection{Self-Efficacy \& Human-Powered Access Technology}
Prior work in HCI has already argued for the importance of self-efficacy in other contexts. \citet{Macvean2013Understanding} highlighted self-efficacy as a key component when designing exergame interventions. \citet{Friedman2018Using} used social self-efficacy to understand the impact of telepresence robots for people with developmental disabilities. Yet, self-efficacy assessments are still not pervasive in HCI; nor in accessibility research. 
People with disabilities have always created support networks to help them deal with accessibility problems, promoting their independence \cite{Bigham2011design}. With technology, and particularly with access to an unlimited number of volunteers/workers, these support networks can be expanded, providing greater coverage and availability. They can be designed to assist with anything, from helping with CAPTCHAS on websites (Webvisum \cite{Webvisum}), to answering visual questions \cite{Bigham2010VizWiz:}, increasing users' control over their technology.  
The emergence of smartphones in our daily lives spurred the creation of novel assistive technologies, many that rely on human computation. Services like BeMyEyes \cite{BeMyEyes2015}, VizWiz \cite{Bigham2010VizWiz:}, Aira \cite{Home} allow users to be connected to others, to assist them by having access to photos or even the camera feed. People can rely on these services for simple tasks, such as color identification, to more subjective (i.e. “do I look good?”) \cite{Brady2013Visual} or complex ones (i.e. teaching a class) \cite{Lee2018Conversations}. However, to be able to leverage them, people have to be proficient, and confident they are able to do it. As a result, many of these technologies are only used by tech savvy individuals \cite{Takagi2009Collaborative}.
There is a need to guarantee that technology is inherently approachable, accessible, and easily leveraged to empower its users. We believe there is an unexplored opportunity to provide pervasive assistance on smartphones through human-powered support. We hypothesize that this assistance can impact the user’s agency and perceptions of self-efficacy over the device.

\section{RISA: Nonvisual Smartphone Assistant}
People need to feel in control of their devices and confident in their abilities to explore and engage in learning processes. We argue that by providing assistance that supports users in their tasks, we will impact their belief on their own abilities. Assistance should therefore be able to adapt accordingly to user interactions, context, be pervasive, and assistive content responsive to users’ requests. A successful assistant must be both competent and trusted \cite{Leporini2012Interacting}.  
To deal with the idiosyncrasies of human behavior, such system is required to be highly flexible, and capable of an understanding beyond current automatic approaches. Building on these requirements and informed by the literature, we developed RISA, a Rich Interactive Smartphone Assistant that leverages human contributors to generate its assistive content. RISA was developed as an Android Accessibility Service  acting as a layer on top of any application, providing task assistance step by step.  The service works in tandem with the Android default screen reader TalkBack.  Herein, we describe our system and report on a user study assessing its ability to support 16 blind people interacting with smartphones.

\subsection{Designing for Rich Interactive Smartphone Assistance}
 In the proposed solution there are two distinct stakeholders: people that contribute with task playthroughs (sighted \textit{Authors}) and end-users who will make use of the task assistant generated by RISA from a \textit{Author} playthrough (blind   \textit{Users}). To avoid the shortcomings of prior works \cite{Rodrigues2019Understanding}, specifically the inability to fully support a task without co-located expert assistance,  we designed RISA to reduce the amount of effort required by \textit{Authors} and leverage as much as possible from available metadata from the smartphone to enrich and adapt assistance. For \textit{Users}, the assistance needed to be pervasive, dependable and unobtrusive.  

\begin{figure} [t!]
\includegraphics[width=\textwidth]{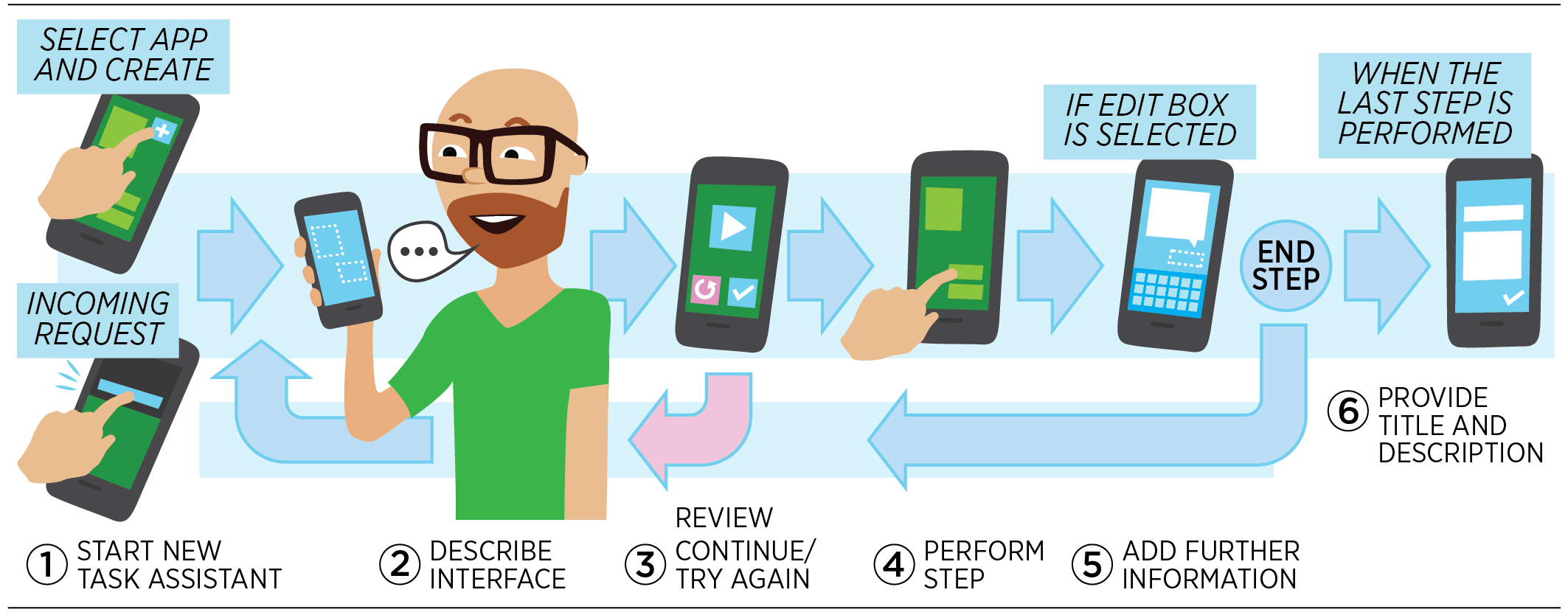}
\Description[RISA authoring scheme with 6 steps.]{ 1) Start new task from an incoming request or by selecting an app from the list, 2) Describe the interface of the screen by audio recording; 3) Review the description and continue or try again; 4) Perform the step as usual by selecting a target; 5) If an edit text box was altered authors are asked to provide additional information via text. After a step demonstration authors indicate whether there are more steps to demonstrate and go to 2) describe interface again; or go to 6) to finish the task by providing a title and a description.}
\caption{RISA authoring procedure.}
\centering
\label{figure:1-risa-authoring}
\end{figure}

 Throughout the design iterations, and informed by our team expertise  and prior work, we focused on \textbf{streamlining the authoring process}, to eliminate the need for \textit{Authors} to have intrinsic knowledge of accessibility and the barriers faced by blind people. We wanted to create a workflow that would enable the \textit{Authors} to simply demonstrate the task as though they were performing it for themselves. To then compensate for the differences in visual and nonvisual interactions (i.e. interacting with a screen reader requires a different set of gestures), and ensure \textbf{consistency in the delivery} of assistance, we capture the interaction logs (e.g. YouTube > Library > History > How to make olive oil > Share) and interface elements from each DOM tree (e.g. element type; bounds; closest text; etc). This information enables us to generate homogeneous instructions despite the possibility of relying on different sources. RISA was designed to rely on \textit{Authors} to demonstrate the task. Then, from the data collected, RISA is able to generate a versatile task assistant. In a following section we detail the authoring procedure (Fig.  ~\ref{figure:1-risa-authoring}).

Task assistant instructions needed to be tailored to each specific device, even the specific location within the screen, as this is highly volatile (e.g. a share button in a scrollable view may not be in the same location). Additionally, \textit{Users} often make navigation mistakes and if the assistance provided does not adapt, it would no longer align with user’s current context. In response, we designed RISA to monitor the \textit{Users}’ interactions and \textbf{adapt the assistive instructions and hints to the current context} (e.g. target location is provided based on the user current DOM tree and not the \textit{Authors}). Monitoring also enabled us to provide path recovery support (e.g. “\textit{You can resume the task from the step History that is currently at the bottom left corner}”  - RISA real-time generated Hint ). After \textit{Authors} create a task, it becomes available to all \textit{Users} (e.g. learn how to share a recently seen YouTube video to Facebook). In a following section we detail how RISA guides step by step (Fig.  ~\ref{figure:2-risa-playthrough}).

\subsection{Authoring Task Assistance}
 \textit{Authors} start by either selecting an app to create a new task assistance or select an incoming request from one of the RISA \textit{Users} (Fig. ~\ref{figure:1-risa-authoring} - {Step 1:} Start new task assistant). RISA now begins capturing the screen information and interaction behaviors (i.e. described in detail in the following section). With each new screen of the app loaded, \textit{Authors} are asked to describe the structure of the active window via voice input (Fig. ~\ref{figure:1-risa-authoring} - {Step 2:}  Describe interface).  While RISA could be developed to automatically describe the interface, it would lack the ability to highlight the most relevant features or to describe the interface in a way that would make sense conceptually with the content. For instance, an author might describe the Google Translate interface as a top part where you write what you want translated and a bottom part where you can check the corresponding translation. RISA would not be able to interlink the purpose with layout, thus we requested this from Authors.  \textit{Authors} are invited to review the recorded description of the screen, if they are happy with the recording (Fig. ~\ref{figure:1-risa-authoring} - Step 3: Review), they are then asked to demonstrate the step (Fig. ~\ref{figure:1-risa-authoring} - Step 4:  Perform Step). 

In steps where text is entered, or potentially sensitive information, or dynamic is identified, \textit{Authors} are prompted to provide additional information. As this content will not match the \textit{User’s } (e.g. when filling a \textit{add a contact} form the name and number to enter will not be the same, thus \textit{Authors} are prompted to provide general indication of what is needed in each field - 'add the contact name for example Brenda')  (Fig. ~\ref{figure:1-risa-authoring} -Step 5: Add Information).  \textit{Authors} repeat this process until they have demonstrated the task in full. At any point they can check the previous steps or save the task. When finished, \textit{Authors} are asked to provide a title and a description for the task (Fig. ~\ref{figure:1-risa-authoring} - Step 6: Title and description); unless this task originated from a \textit{User} request (as it would already have both).

\subsubsection{Data Collection} 
 We rely on the authoring process to collect a mix of metadata and interaction data. We developed RISA as an Android Accessibility service which  enables us to collect the following information for each screen the \textit{Author} goes through (Fig. ~\ref{figure:1-risa-authoring} - Step 4:  Perform Step): clicked view (view data non-exhaustive list - activity, bounds, available action, class name, closest text, text, content description, dimensions, type, timestamp, visibility status), interactive views (view data from each),  audio file of the interface description given by the \textit{Author}, application activity and package name, scrolled views (view data for views that are scrollable, are were scrolled during authoring), timestamp, screen title and type of interaction performed (e.g. click, long click). RISA, as any other service, operates at a system-wide level without any knowledge of the logic implemented by each application nor the ability to uniquely, and consistently, identify views (i.e. content).
Therefore, RISA relies on textual descriptions of elements to identify targets and to guide \textit{Users}, thus descriptions are essential. For every view, we locate the closest text or content description, first through its hierarchical children, then its parents.

\begin{figure} [t!]
\includegraphics[width=\textwidth]{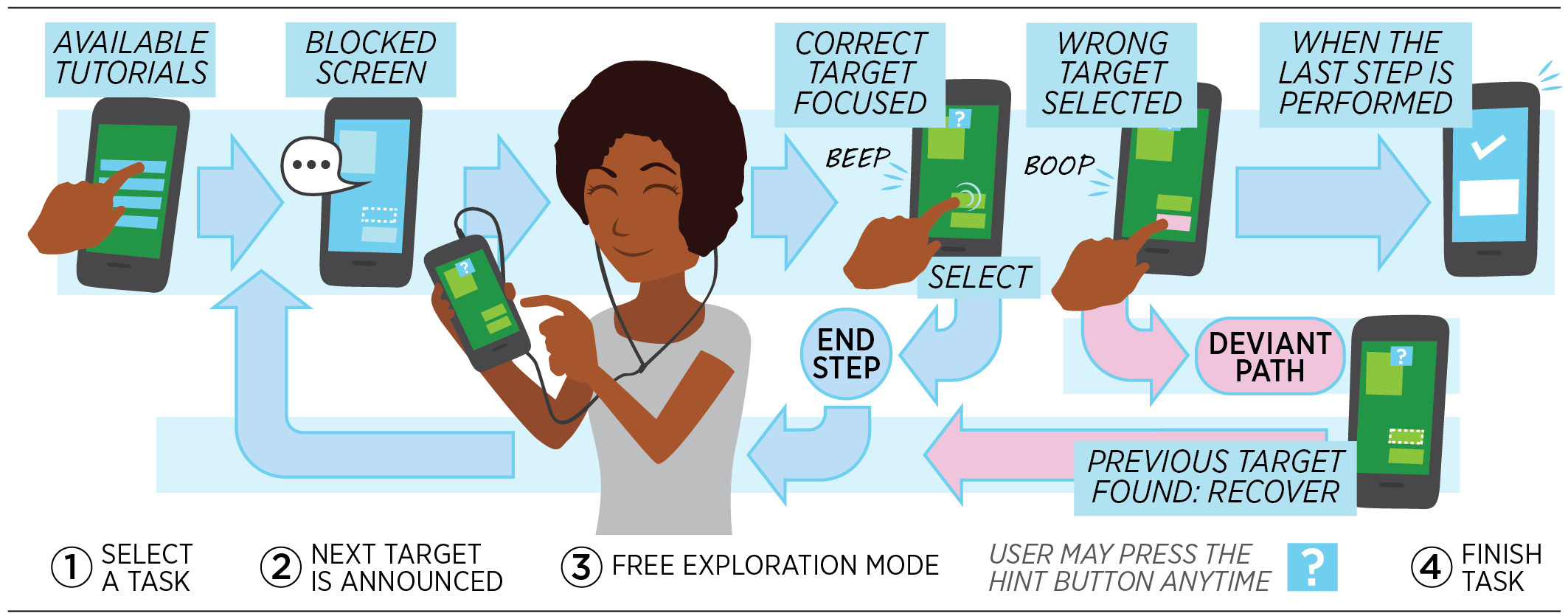}
\Description[RISA playthrough scheme with 4 steps.]{ 1) Select task, 2) Next target is announced with the screen blocked, people listen to the target description 3) Users enter free exploration mode. When the correct target is focused users receive a beep. If they select it they either go to the next step 2), or it was the last step they go to 4) finish task. During free exploration, if people select the wrong target they hear a \textit{boop} and RISA enters recovery mode. If RISA detects a previous target on screen it resumes the task on that target moving to 2) next target. In 3) free exploration, at any time people may select a hint option to receive an hint according to the task and monitored behavior, details in a following section.  In 4) 'finish task' hear a audio tune symbolising success. }
\caption{RISA playthrough procedure.}
\centering
\label{figure:2-risa-playthrough}
\end{figure}

\subsection{Using the Task Assistant}
RISA \textit{Users} can browse through each of their installed apps and check which ones have supported tasks, check their title, description, and steps (Fig. ~\ref{figure:2-risa-playthrough} - Step 1: Select Task). If a task is not available, \textit{Users} can create a request describing what they are looking for through text, or by sending a voice message associated with a particular app. When \textit{Users} select a task, RISA initializes the app on its first screen, and begins guiding step by step. During the task, RISA behaves somewhere in between an assistant, that only intervenes when requested, and an interactive tutorial (detailed in the following section). It relies on the data collected during the authoring process to generate each of its instructions, and hints, through the device Text to Speech Engine (TTS), ensuring instructions delivery consistency between different \textit{Authors}, and different applications.

\subsubsection{Matching Targets}
We rely on a combination of the views’ characteristics, and the information provided by Authors (i.e. view is dynamic/sensitive) to match target views with current screen views. Target location is never used for matching purposes as it may differ from device to device, screen orientation, app version, time of day and many other factors. When locating a target on the screen, RISA checks all the interactive elements (i.e. clickable, long clickable) that are visible on screen. If it finds a target that shares the same package name, class name and closest text, it assumes it has found a match. If multiple matches are found, RISA compares element sizes, then hierarchical level (i.e. depth of each element in its DOM tree); if matches are still equal, then it selects the first element to appear on the DOM tree. 

\subsubsection{Guiding Step by Step}
RISA always follows the same procedure. First, it blocks all screen interactions with an overlay (i.e. with the exception of the home and back button) while it announces the description of the next target element (Fig. ~\ref{figure:2-risa-playthrough} - Step 2: Target Announcement). Then, \textit{Users} enter a free exploration mode, where they interact with the app as they want, without interaction interference from RISA (i.e. RISA does not prevent or makes any interaction). During free exploration (Fig. ~\ref{figure:2-risa-playthrough} - Step 3:  Free Exploration), RISA creates an overlay at the top center of the screen, with a \textit{Hint} and a \textit{More Options} element; and monitors \textit{Users}’ interactions. The top center of the screen was chosen as the default position because, typically, there are no interactive elements in the top center, and when there are, they usually occupy a larger area than RISA covers. Moreover, since it is directly underneath the typical place for smartphone speakers it has a clear tactile cue for quick access.

\textit{Hints} can be selected at any time, and when focused, RISA blocks again the interaction and provides the next hint (i.e. using the Text-to-Speech engine it announces the generated hint)  depending on the target, and the interactions monitored (described in detail in a following section). Through\textit{ More Options}, \textit{Users} can control the task flow (i.e. exit, restart).

During free exploration, if \textit{Users} focus the correct target element, RISA provides an audio cue (i.e. a short ‘beep’). If \textit{Users} select anything but the correct target, RISA gives an audio cue as a warning (i.e. a short ‘bop’). To support path recovery, when \textit{Users} deviate from the path, RISA checks the current screen for any view that matches a previous step target. If it finds one, it will initiate the default procedure.

When a correct target is selected, RISA blocks interactions again and provides the next target description. RISA repeats this behavior until the last step is performed, after which it provides an audio message stating the task was successfully completed, and a short audio clip symbolizing success (Fig. ~\ref{figure:2-risa-playthrough} - {Step 4:}  Finished).

\subsubsection{Text to Speech}
RISA uses the Text to Speech (TTS) engine available on the device for all feedback, and selects a different voice from the screen reader to ensure intelligibility of speech \cite{Brungart2005Improving}, facilitating \textit{Users}' ability to distinguish between RISA and the screen reader. The only exception are the interface descriptions recorded by \textit{Authors}, which are provided in their original form.  Since Android 8.0 the accessibility volume can be controlled independently of the multimedia volume. RISA uses the multimedia API to play instructions, thus it does not interfere with the default behaviors, and feedback of the screen reader. 

\subsubsection{Target Announcement}
RISA default target announcement behavior is to state the required interaction, followed by the target closest text (e.g. “Select Create Contact”). 

If, during authoring, text was written in any form edit text box, first RISA announces how many edit boxes will have to be edited, then what element to select afterwards, followed by an enumeration of what to write in each of the edit text boxes (e.g. contact name, number, email). 

\subsubsection{Additional Hints}
When the \textit{Hint} element on the top center is focused, RISA plays a single hint from the next hint type available (ordered below). RISA default behavior is to play the next hint, from A to I. If a hint is not available or does not make sense given the target or previous monitored interaction (details below), then it skips to the next one. When it reaches I) Recover, it circles back to A) EditBox Detail. The flexibility of the provided hints is fundamental to be able to correctly assist the user depending on its current context and target. Hints are adapted depending on the user path, if and how far it deviated, and how much information RISA gathered during the authoring procedure. Below we describe how the hint types behave.

\textbf{A) EditBox Detail}, the first hint type, is only available if there were EditBoxes edited during this step in the authoring process. The hint uses the text provided by the \textit{Author} when prompted (e.g.”\textit{What information is required in this text field? – The new contact name}”). It is also the only hint type where RISA will not skip to the next type until it has provided all hints of its type (e.g. \textit{“First write the new contact name}”). 

\textbf{B) Long Click Instruction} is only available if the element was long pressed. It describes the touch interactions required to do a long press with a screen reader.

\textbf{C) Target Position } is available if RISA can find the target on screen; thus, for targets tagged as sensitive or dynamic it is not usually available. RISA searches based on textual descriptions (i.e. text, closest text, content description), and only using dimensions and next, hierarchical position on the ‘DOM tree’ when multiple matches are found. When a match is found, its absolute location is retrieved. Target location was divided into nine sectors: corners, edges and center, since corners and edges have been shown to be preferred by visually impaired people \cite{Kane2011Usable}. Since larger targets can be in multiple locations, the labeling priority is always given from corners to edges, to center, from left to right and from top to bottom (e.g. an element that occupies the full top edge will provide a hint stating “\textit{top left corner}”).

\textbf{D) Scrollable} is available if the user has not deviated from the path, if the target is not on screen, and if the target is known to be in a scrollable element. During the authoring process RISA collects the closest text to the scrollable view (i.e. capture in the list of interactive views) where the target element is. RISA then uses this text to try and find a representative element of the scrollable list, if it does then it announces the hint (i.e. navigate in the list that has X).

\textbf{E) Swipe Scroll}, available under the same restrictions as D) Scrollable. It describes the touch interactions required to interact with the identified scrollable element (i.e. “\textit{Find}” + <closest scrollable view text> +  “\textit{and navigate the list by swiping from left to right until you find the target}”).

\textbf{F) Target Not Found} is available if the target is not on screen, is not inside a scrollable element, and the user has deviated from the path. This can notify the user to look elsewhere instead of spending significant time going through all existing interactable elements.

\textbf{G) Layout Description} plays audio recording of the interface description given by \textit{Authors}. Since RISA does not impose any limit to the length of the layout description, for this hint only, if \textit{Users} double tap on the blocking overlay it stops the hint and resumes free exploration.

\textbf{H) Repeat}, it repeats the target announcement. If the \textit{User} has deviated from the path, it checks to see if any previous target is on screen. If it is, it makes the target announcement for it as per standard behavior.

\textbf{I) Recover}, is only available if the \textit{User} has deviated from the path and it ensures \textit{Users} are aware, they can resume the task from any step.

\section{Study Design}
To assess the impact of effective assistant on users’ self-efficacy, we conducted a comparative user study where participants performed tasks with and without the assistance of RISA.  First, still during the development of RISA, we conducted two informal sessions with each stakeholder (i.e. an \textit{Author} - recruited at the university and a \textit{User} - recruited from a local institution) where they freely explored RISA and provided additional feedback which resulted in adjustments to verbosity and prompts descriptions. The informal sessions served as a first deployment of RISA to ensure the study protocol and prototype were comprehensive and robust. These users were not included in the ensuing study.

Next, to create the content to be used in the study, we conducted an authoring session with sighted users with no accessibility knowledge, where participants were asked to demonstrate a set of tasks. Next, to assess the effectiveness and the impact of the assistant, we recruited 16 blind people to participate in two playthrough sessions. Except for tech savvy users, people will often have trouble describing their issues \cite{Bigham2017Effects, Takagi2009Collaborative} and report to rely on others for assistance or have no assistance at all. We intentionally compared our system against the other current available independent alternative, that most report to rely on, which is exploring by themselves.  Therefore, in one condition, participants performed a set of tasks with no assistance, as they would do on their own device. In the other, participants performed a set of tasks with the help of RISA. 

\begin{table}[]
\resizebox{\textwidth}{!}{%
\begin{tabular}{llll}
\hline
ID  & CD & App              & Task                                                                                                                              \\ \hline
tt1 & -  & Contacts         & Add new contact                                                                                                                   \\
tt2 & -  & Contacts         & Add \textless{}X\textgreater to favorites                                                                                         \\
t1  & A  & YouTube          & Find a video from History,   share to Facebook                                                                                    \\
t2  & A  & Netflix          & Check in-app notifications,   find a series. Add it to your watch list and download an episode.                                   \\
t3  & B  & Uber Eats        & Order a Big Mac, medium   size with coca-cola from McDonalds \textless{}location\textgreater{}. Check-out.                        \\
t4  & A  & Google Translate & Switch translation   from X-Y to X-Z. Listen to previously stored \textless{}phrase\textgreater{}. Play it in Z.                  \\
t5  & B  & One Football     & Check followed   competitions. Check the teams in \textless{}country\textgreater league. Check \textless{}team\textgreater squad. \\
t6  & B  & Outlook          & Turn on the ‘Do not   disturb” mode.                                                                                              \\ \hline
\end{tabular}%
}

\caption{Task condition (Cd) and description, with details removed for blind review.}
\label{tab:task-table}
\end{table}

\subsection{App \& Task Design}
Authoring and playthrough participants were asked to do eight tasks (T). The same two were always used as training tasks (TT) (Table \ref{tab:task-table}).  RISA was designed to be able to handle the complexity and variety of mobile applications as a third-party service. To highlight its capabilities and to verify its adaptability, we selected six mainstream applications from Google Play Store, from six different categories and designed a task for each one (Table \ref{tab:task-table}). Moreover, by relying on mainstream apps instead of custom-made applications for laboratory assessments, we are one step closer to an ecological valid environment. 

We chose one of the top apps of each of the selected categories (Google Play Store -  Portugal  ranking) that had enough complexity to create a task with more than three steps and basic accessibility compliance (i.e. not having every element unlabelled) – this was verified by the research team using the TalkBack service. For the category \textit{Music \& Audio}, although \textit{YouTube Music} was the top app we choose its more pervasive sister app, \textit{YouTube}, that does not appear on top categories due to being pre-installed in most devices. 

Tasks ranged in number of required steps and workflow structure. The tasks T2 and T6 had a single sequence of steps that had to be performed with no alternative paths available, and never passing through the same screen. The remaining tasks had alternatives paths available in some of its required steps (e.g. multiple ways to reach the restaurant). In T3 and T4 participants had to pass through the same screen at least twice to complete it.

In the authoring session tasks were counterbalanced. For the playthrough session the tasks were randomly assigned to group A and B (Table \ref{tab:task-table}). Within a group of tasks participants performed them always in the same order. For group A the order was randomly generated as T1, T4, T2; and for B T6, T5 and T3. The group of tasks was counterbalanced between the two conditions (with RISA and without). Participants performed both conditions, one in each session (i.e. conducted on different days). The first assigned condition was also counterbalanced between participants.

\subsection{Apparatus}
We used a Xiaomi Mi A2 device running Android 8.1. For every app with a task, a shortcut was made available on the device home screen. RISA was pre-installed on the device. In the playthrough sessions we used TalkBack, the default Android screen reader. Participants were requested to use headphones for the playthrough sessions.

\subsection{Authoring Session}
For the authoring phase we recruited 13 sighted participants with no or limited accessibility knowledge, only one had previously tried a desktop screen reader and none knew how blind users interacted with smartphones. Ages ranged between 19 and 52 (M=26.77, SD=8.80), two iOS users and 11 Android. In the following sections we will refer to authoring participants as \textit{Authors}.

We informed participants they would be demonstrating how to do a set of tasks while using the RISA authoring feature, and that their demonstration would be used in the following weeks to guide blind people. First, participants were given a brief overview of the RISA authoring features, then they were guided through the authoring process for TT1. Next participants demonstrate TT2 without any guidance but were encouraged to ask any questions they had. After the training tasks, participants were asked to demonstrate 6 tasks (Table \ref{tab:task-table}). Before recording each demonstration, participants were guided through the task ensuring they were familiar with it. They were encouraged to explore and ask any questions.  When they felt comfortable with it, they started the RISA authoring process.  \textit{Authors} were rewarded for their time with a gift card.

\textit{Authors} were able to successfully provide 78 demonstrations, 13 of each task. From those we discarded 11 due to technical issues with the recordings of the interface descriptions.  All \textit{Authors} were able to use RISA authoring capabilities without any assistance, assessing the authoring process on average 4.85 when asked to rate how easy it was to complete the tasks (i.e. on a scale from 1 to 6 from extremely difficult to extremely easy).  Every \textit{Author} had at least three of their demonstrations followed by blind participants. 
\subsection{Participants}
For the playthrough sessions we recruited 16 legally blind participants , from local social institutions, ages ranging between 26 and 71 (M=46.63, SD=12.64), 9 iOS users and 7 Android, with a variety of self-reported smartphone expertise (M=3.56, SD=0.79) (i.e. participants were asked to rate their smartphone expertise on a scale from 1 to 5 where higher is better), with experience between 4 months and 7 years,  all had previous experiences with Android devices,  and all relied on the smartphone screen reader . Participants were required to be able to at least make a phone call and send a text message with their smartphone. 

\subsection{Procedure}
Participants were informed that the purpose of the study was to understand what impact a smartphone task assistant could have. After a demographics and smartphone expertise questionnaire, participants were asked about their familiarity with the applications and tasks chosen for this study. Only two participants had previously shared a YouTube video on Facebook but not from History, and one participant had ordered food on Uber Eats with her iPhone. Each participant was informed they would be performing a set of tasks in two conditions: 1) with the assistance of RISA; 2) as if they were home trying to perform the task by themselves. Each condition was counterbalanced and performed in different sessions, in different days, each lasting about 1 hour.  Each participant was randomly assigned to start with condition 1) or 2) for their first session. 

In both, participants were first asked to find the contact John to get used to the device. Then participants performed two training tasks (i.e. TT1 and TT2). In 1) with RISA, TT1 was used to present to participants the playthrough features of the assistant. Participants were guided through the task, encouraged to consult hints, instruct on the type of hints and audio cues provided, the assistant guidance behavior and the recovery mechanisms available. For TT2 with RISA, participants were asked to follow the assistant alone, and encouraged to ask any question they had. After participants finished TT2 they proceeded to do set A or B of 3 tasks (Table \ref{tab:task-table}) in one of the conditions. Tasks within a set were always presented in the same order. Participants performed a different set in each condition. We ensured each set was performed the same amount of times in each condition. In the condition with RISA we randomly selected one of the demonstrations created for that task, ensuring that each participant followed demonstrations by three different \textit{Authors}. No pair task-author was ever repeated. 

Participants were asked to complete the task at hand to the best of their efforts. Each task was introduced as described in (Table \ref{tab:task-table}), “In application X do Y”, participants started from the first screen of the application. In 2) (without RISA) participants were told to notify the researcher when they had completed the task, the researcher also observed and took notes; in 1) once they had performed the final step, RISA announced the end of the task. 

If participants did not make any progress toward the task after 5 minutes, the task was interrupted and considered to be unsuccessful. If participants alerted the researcher that they were not able to complete the task, or showed visible signs of frustration, the task was also interrupted. Participants could ask the researcher to repeat the task description at any time, but no additional help was provided. At the end of each condition (i.e. in different days), each participant filled a brief questionnaire about smartphone self-efficacy based on Bandura’s \cite{Bandura2006Guide} work, with    a 100-point confidence scale from 0 (“Cannot do”) to 100 (“Highly certain can do”). The 11 questions were informed by prior work on computer-self efficacy\cite{Compeau1995Computer} and smartphone self-efficacy \cite{Hong2014Using} scales, but adapted to the specific context based on challenges identified in previous work \cite{Bigham2017Effects, Grussenmeyer2017Accessible, Rodrigues2015Getting} (e.g. learn a new App, understand when something does not work) as state by \citet{Bandura2006Guide} \textit{“scales of perceived self-efficacy must be tailored to the particular domain of functioning that is the object of interest}”.

In the debriefing of the second session we conducted a semi-structured interview to understand user confidence with and without RISA, perceived accessibility and self-efficacy. Participants were encouraged to share their thoughts on learning, exploring, assistance and about RISA. Participants were rewarded for their time with a gift card. 

\subsection{Design \& Analysis}
To understand the effect of assistance we must first ensure the assistance provided was successful. Thus, in the following section we report the impact of the condition on task and participant success rate. A task was considered to be successful if the participant performed all the required steps, independently of the order. For example, if for the Uber Eats task, the participant did all the steps but ordered from a different McDonalds than the one requested, the task would be considered unsuccessful. We used task success rate as the dependent variable and relied on a mixed effects model analysis following the procedure in \citet{Seltman2012Experimental}. We modeled Condition and Task as fixed effects to ensure we account for the possible effects of the differences in task complexity. Task and participant were also added as random effect to accommodate repeated measures. Task (p=0.879) and Condition*Task (p=0.815) were not significant fixed effects, thus, to simplify the model we only accounted for the fixed effect of Condition (p<0.001).

We transcribed all interviews and conducted a qualitative analysis using, primarily, a deductive coding approach. Two researchers independently coded one interview to revise the initial codebook, adding two new codes and changing other two. Then the two researchers coded other four interviews independently. We calculated a Cohen’s kappa agreement of k=0.66 (SD=0.32), which represents a fair to good agreement. One researcher completed and revised the remaining interviews. We were interested in understanding self-efficacy, perspectives on accessibility, confidence when using smartphones and whether, or not they were affected by having effective pervasive assistance. We further support our qualitative analysis with the results of the questionnaires on self-efficacy. Considering a within subject design, and since self-efficacy was normally distributed, we applied a paired samples t-test.

\subsection{Findings}
In this section, we analyze results of the playthrough session. When we refer to participants, we are specifically referring to those that participated in the playthrough session. We highlight their perspectives around the task assistant and its impact, which represents the basis for our discussion around self-efficacy. As presented in our motivation, participants discussed the need to permanently learn new apps and features.

\begin{quote}
\textit{“Everything is always changing... The systems are always updating, it is always better to learn new stuff.” --P3 \\}
\end{quote}

\begin{figure}
\centering
\begin{minipage}{.5\textwidth}
  \centering
  \includegraphics[width=0.95\linewidth]{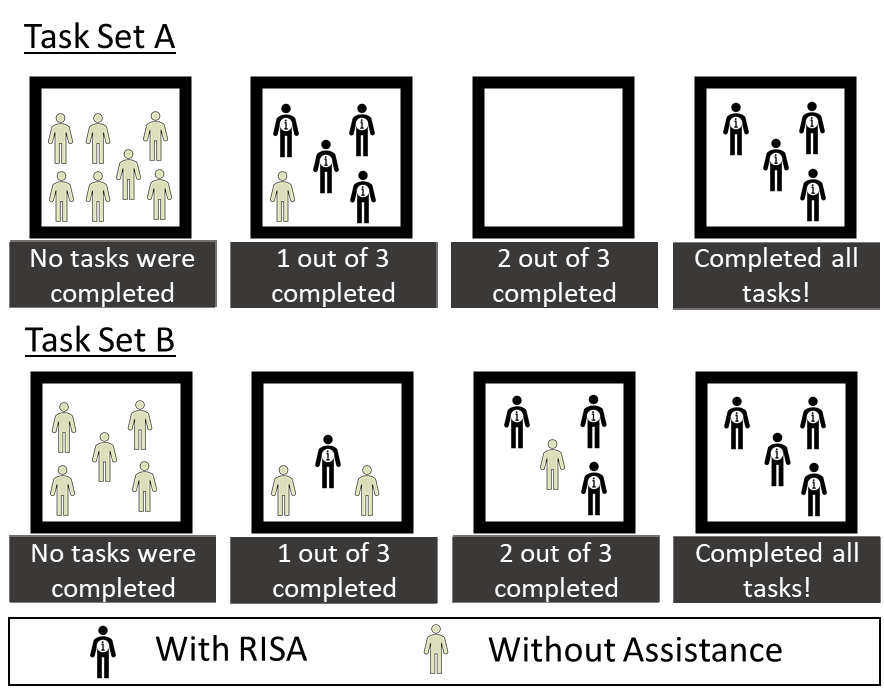}
  \Description[Number of tasks successfully completed by condition with all (16) participants with RISA completing at least one task. Only 4 participants without RISA completed one ore more tasks]{ For each task set, four boxes are displayed with zero, one, two or three tasks successfully completed.  For task set A, seven participants without RISA completed zero tasks, and one completed one. Four participants with  RISA completed one, and other four completed all tasks. For task set B, five participants without RISA completed no tasks, two completed one, and one completed two. One participant with RISA completed one task, three completed 2 and four participants with RISA completed all tasks.}
  \captionof{figure}{Number of tasks successfully completed for \\ task set A and B for each condition.}
  \label{fig:fig_success}
\end{minipage}%
\begin{minipage}{.5\textwidth}
  \vspace*{0.9cm}

  \includegraphics[width=1\linewidth]{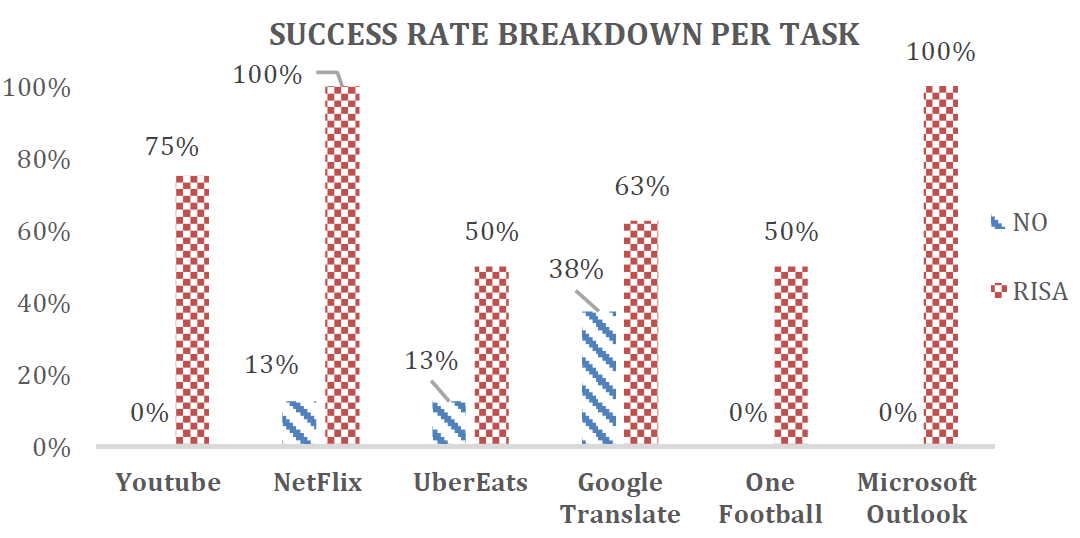}
    \vspace*{0.40cm}
      \Description[Bar graph with all applications showing RISA with a higher  success rate superior than without. ]{Netflix, and Outlook RISA has 100\% success rate against, 13\% and 0\% respectively. The closest values are in google translate with RISA 63\% and without 38\%.}
  \captionof{figure}{Success rate breakdown per app between the condition without RISA (NO) and with RISA.}
  \label{fig:fig_graph_success}
\end{minipage}
\end{figure}

\textbf{Success Rate.} Answering RQ1 - when using RISA, participants were more successful. The condition with RISA had an average success rate of 72.9\% (SD=0.45) against 10.4\% (SD=0.31) without. The estimates of the fixed effects calculated using a mixed model, shows that condition (F=64.849, p<0.001) had a significant effect on participants’ success rate. The condition without RISA had a negative effect on the task success rate, between 76.9\% and 46.5\%.

In Figure ~\ref{fig:fig_success}, we can observe the success of the two different task sets, per condition. Half of the participants were able to complete all tasks, and all completed at least one of the three tasks successfully with RISA. 

On the other hand, without RISA, the majority of participants (12) were unable to successfully complete a single task. When we analyze the success rate per task (Figure ~\ref{fig:fig_graph_success}), the data suggests that, for some tasks, RISA was highly effective (e.g., Netflix and Outlook), while for others we only observed marginal improvements (e.g., Google Translate).  For Google Translate, although RISA provided an audio cue when the correct element was focused, it was not enough to overcome the duplicate element with a different function that was on screen. The wrong one was first on sequential navigation, and close to an edge and was therefore easier to reach.\\

\textbf{Hints Consulted.} With RISA, participants were able to consult hints at any time during exploration. A total of 300 hints with an average of M=6.00 (SD=5.77) per task, and   M=6.14 (SD=4.72) per participant were triggered during playthrough sessions. In 20.8\% (10) tasks, participants did not rely on any hints, and only followed the step-by-step  instructions. Of those, in 8.3\% (4) tasks, participants were unsuccessful. As expected, participants had a wide variety of hint consulting behaviors, ranging from individuals who consulted a single hint during the three tasks and only completed two tasks successfully, to one that consulted 60 and completed all tasks successfully with RISA. Hints required users to actively seek out assistance. In this study, some participants leveraged them to their benefit, highlighting the importance of having additional adapted assistance to complement step by step target instructions.\\

\textbf{Structure based learning.} When learning a new app, particularly when out of curiosity, participants reported how it is a slow and methodical procedure of trying to make sense of the underlying interface. All participants described how first it is all about exploring the screen, to try and create a mental model that allows them to interact successfully. The amount of effort it takes is highly affected by user expertise and app accessibility. Only users who are already confident in their ability to explore will engage with applications in this manner.

\begin{quote}
\textit{“Some apps are complex. It takes a while before I figure out what I can do, how to navigate, it's complicated.” --P6}
\end{quote}

RISA was effective in providing assistance but was not perceived to support learning through exploration, but rather as an alternative. For expert users who are comfortable learning through exploration, RISA current support, might not be the adequate solution.

\begin{quote}
\textit{“(RISA) is important when we want to complete a specific task, but for exploration I cannot tell if it is functional, if it helps. RISA gives accurate references, what happens is I focus on them and completely ignore the rest.” --P15\\}
\end{quote}

\textbf{Task based learning.} The alternative described by participants to exploration was task-based learning. When learning something out of necessity, when time was a factor, participants reported how they are simply trying to figure out how to do a task ignoring all the rest. Additionally, participants described their learning process as task oriented once they have a basic understanding of the app. RISA was associated with this kind of learning by all participants.

\begin{quote}\textit{“I always go through the options it has, then I try to accomplish my goal (...), if I can’t I ask for help (...), then with repetition I am able to do it...” -P11 }\end{quote}

Participants believed RISA would enable them, and others to be more efficient when trying to complete a new task. For example, P6 state \textit{“(RISA) helps getting where we want to go faster. With efficacy and without making mistakes.” P6} RISA was seen as a way to prevent early negative experiences and quickly empower novices with a sense of control.

\begin{quote}\textit{“I believe it's quite a useful tool, particularly for those who are starting to explore smartphones. It ends up guiding the user faster, more efficiently to the right places where one should press.” P16} \end{quote}

Additionally, 13 participants believed they would be faster with RISA, affecting their feelings of competence, in this case estimated speed.
 \begin{quote}\textit{“Almost everyone, even for more experienced people. Even when they learn a new application by instinct, they would get there faster. I would recommend even for experienced people.” --P11\\} \end{quote}
\textbf{Applications’ perceived accessibility.} Participants reported their self-efficacy to be influenced by how they perceive the app's accessibility. When first exploring an app, if basic accessibility guidelines are not followed (e.g., buttons have labels), participants can quickly dismiss the app as inaccessible and give up.
 \begin{quote}
\textit{“It is always a challenge to understand if an app is accessible or not. If not, it is enough to make me give up on it.” P15}

\textit{“It all depends on how an app is built. My confidence is proportional to the degree of intuitiveness of the apps features.” P16} \end{quote}

Two of the tasks had inaccessible elements: Netflix had elements with no labels; Uber Eats had labels in a foreign language unknown to most of the participants, especially when we consider a screen reader trying to pronounce in a different language. One of the major struggles we observed during the trials for these tasks was dealing with this content when navigating. When using RISA, participants quickly dismiss inaccessible content, or not, based on RISA guidance. When using RISA participants had additional information about the structure, what to look for, and audio cues. Thus, even when the target was inaccessible (e.g. no label), they would be able to determine to select or move on to explore the rest of the screen. Conversely, without RISA, as soon as participants were faced with inaccessible content they would start wondering what was happening, or if they had to engage with that particular content.  Participants associated RISA with being helpful/useful and responsible for making tasks easier. Despite not making content accessible, RISA gave users the ability to overcome inaccessible content on their own. Effective assistance appears to influence the perceived complexity of the task.

\begin{quote}\textit{“Without any room for doubt, today everything was simpler. The options were not always obvious, and with the assistant I knew exactly what to look for.” --P17\\}\end{quote}

\textbf{Self-efficacy and effective assistance.} Participants expressed they are not confident when trying new apps. People are often left with no choice other than trying even if they are not confident.

\begin{quote}\textit{“A bit afraid of messing up, that I block it. Deleting stuff that I shouldn't or installing something that I have to pay and not understanding it.” --P3}

\textit{“Not confident, not really. However, I try to understand what is going to happen.” --P2} \end{quote}

RISA was thought of as a tool that would: 1) facilitate the learning process; 2) enable doing tasks faster; 3) provide clear paths thus avoiding doubts and fears; and 4) facilitate the understanding of interfaces. After using RISA, participants reported higher rates of confidence, particularly in their ability to understand their current screen and the overall flow of the application.

\begin{quote}\textit{“For me, it’s a bit difficult to explore new applications (...) (With RISA) maybe I would be more confident.” --P4}\end{quote}

We found that RISA affected the participants ratings of self-efficacy. The average score of the sum for self-efficacy M=91.31 (SD=14.23) with RISA and M=78.75 (SD=19.91) without (i.e. higher scores represent higher efficacy). The paired-samples t-test revealed a statistically significant difference between with RISA and without, t(15)=3.11, p=0.007 (two-tailed), with a large effect size (eta squared=0.39) \cite{Cohen2013Statistical}. 

The confidence reported, together with the findings of the \textit{Task} and \textit{Structure base learning} sections allow us to state that effective smartphone assistant positively impacts user’s self-efficacy (RQ2). 

\section{Discussion}
We conducted a comparative user study to assess how effective nonvisual human-powered assistance impacted users’ perceptions of accessibility and self-efficacy. Furthermore, we showed we can rely on untrained individuals with limited accessibility knowledge to create effective smartphone assistance. Herein, we discuss our findings that should be of interest to researchers and practitioners working on nonvisual mobile accessibility. 

\subsection{Shift in learning practices}
Currently, when engaging with new smartphone applications, blind people have no choice but to conduct extensive exploration of the whole screen to create a mental model of the available options. This procedure is similar to how sighted people engage with new apps by quickly grasping the interface through all its visual cues. For a subset of tech savvy users who enjoy dedicating their time to exploring a whole app before being able to use it the status quo is sufficient. However, for all others the procedure is long and has a substantial workload for complex or out of the ordinary applications.   
Learning how to interact with technology seems to resemble how one learns a second language.

Learning how to first perform gestures, and then learn through exploration by first creating mental models of the structure of an application appears to be the equivalent of the focus on language structure and syntax. Since the 1970’s the assumption has been that focusing only on language structure is not enough, but rather there needs to be associated with an ability to express meaning \cite{Prabhu1987Second, Skehan2003Task-based, Widdowson1978Teaching}; the smartphone equivalent would be the ability to perform tasks. 

Since then, task-based teaching \cite{Prabhu1987Second} has been the prevalent construct by which teachers have created their syllabus \cite{Seedhouse1999Task-based} and the ‘dominant paradigm in the teacher education literature’ \cite{Lynch2000Exploring}. We argue that technology is still far behind current teaching practices, and for blind people the status quo is a decades old approach that forces form before function.

One of the arguments against task-based learning in second language learning is the difficulty in creating tasks that are representative of the real world \cite{Lynch2000Exploring, Seedhouse1999Task-based}. However, the limitations of transposing real-world, useful tasks to the classroom does not apply to technology learning or assistance, as tasks can easily be performed in a real context. We argue that a task-based approach with a focus on meaningful tasks can have a positive effect in a user a sense of self-efficacy that a focus on structure (i.e. gesture, or interface) cannot have. Similarly, past research has highlighted how the ability to perform a task, particularly while adopting a device,  can have profound changes in users’ attitude \cite{Rodrigues2015Getting}.

RISA has allowed to shift the focus from form to function by allowing users to learn in a task-based environment. RISA was not thought of as an assistant that was able to support a methodical exploration of structure, but rather as an alternative. Particularly for novice users, it was described as an easier way to get familiar with applications by performing tasks, while ignoring the rest that would create additional workload. With such assistance, users reported they would be able to focus on tasks, learn, and perform them quicker. After being familiar with a particular set of tasks, users believe it would be easier to understand the application, given they would already be aware of some of the workflows and interaction behaviors. Similarly to learning a second language, once learners are proficient, a task base scenario might only help for unknown contexts.

\subsection{Non-specialists were able to provide effective assistance}
Previous work found a mismatch between the assistant provided by non-specialists and the required by end-users \cite{Rodrigues2019Understanding}. With RISA, we were able to fill this gap by structuring the authoring process, implicitly collecting app and interaction data during authoring; and by, during assistance, monitoring users’ interactions and assessing app structure. Untrained sighted individuals created assistive content that had a positive impact on participants’ task success rate when compared against no assistance. RISA had a positive impact on success rate, going from 10\% without assistance to 70\%.  Playthrough participants completed six different tasks in six different applications from six different categories. Our findings demonstrate RISA is flexible supporting a variety of application contexts and interfaces. We believe there is an opportunity to further explore pervasive system-wide  assistance that acts in between the user and its applications providing support based on implicit contributions by other users. In doing so, we will be preemptively increasing the content that could be leveraged by services, such as RISA, to be able to provide just-in-time learning, further addressing the fundamental issue of the availability of support.  

\subsection{Self-efficacy was affected by effective assistance}
Self-efficacy has been reported to have a positive effect on the decision to use a web technology and influencing actual use \cite{Yi2003Predicting}. Assessing if a solution is effective or not is not enough to assess its potential impact. In this study, we go beyond traditional metrics and make the first attempt to assess blind people self-reported smartphone self-efficacy.

With RISA, participants reported to be more confident in their ability to understand interfaces and apps. RISA use of clear targets enabled participants to ignore all possible confounding and inaccessible elements. We also found participants attributed the improved success rate to an improvement in efficiency. 

Effective assistance appears to have an impact beyond success rate. Our results suggest an effect on how users measure their own ability to perform tasks, how they perceived apps accessibility and complexity. In turn, the increase in self-efficacy may be what is affecting participants perceptions of task difficulty. Similarly, \citet{Agarwal2000Time} showed a strong relationship between application specific self-efficacy and ease of use.

Unlike other assistive technologies that rely on the crowd in real time such as BeMyEyes and AIRA, RISA aggregates knowledge and does not always require real time communication. The asynchrony of communication and the availability of knowledge appears to have affected how people perceive how they accomplish the task. Independence and autonomy were reinforced by always-available support, which seem to have contributed to increased perceptions of self-efficacy. 

Pervasive assistant has the potential to impact user’s overall smartphone use and exploration behaviors. In this study, we observed an impact on how participants reacted to inaccessible content and complexity, and on self-reported confidence to control the device.  We believe the results are promising revealing how effective assistance can support users to learn and explore independently.  

\section{Conclusion}
Despite smartphones being considered 'accessible' to blind people, many are restricted to engage with a mere 'handful' of applications and device features for fear of failure and doubt in their own abilities to overcome barriers or challenges. We propose the need for solutions that support and enrich the more nuanced experiences of using the smartphone and serve to nurture  user’s  self-efficacy and confidence in their own ability to control and use technology. To that end, we developed RISA, a rich interactive smartphone assistant that makes use of human-powered accessibility approaches to create pervasive task assistance for blind and visually impaired people. Our user study involving 16 blind people demonstrated that RISA was not only effective in supporting blind people learning new smartphone tasks, but more importantly improved their perceptions of self-efficacy and confidence with the device. As innovation in consumer technology progresses and enables new digital capabilities, we must ensure to provide assistance to mitigate the effects of the digital 'literacy' divide and support universal adoption.

\begin{acks}
We thank Funda\c{c}\~ao Raquel and Martin Sain in Lisbon (Portugal) and all participants. This work was partially supported by  Funda\c{c}\~ao  para a Ci\^encia e Tecnologia (FCT) through scholarship SFRH/BD/103935/2014, project mIDR (AAC 02/SAICT/-
2017, project 30347, cofunded by COMPETE/FEDER/FNR), and LASIGE Research Unit, ref. UIDB/00408/2020 and ref. UIDP/00408/2020.
\end{acks}

\bibliographystyle{ACM-Reference-Format}
\bibliography{ref-extracts}

\appendix

\end{document}